\documentclass{xsurv_article}

\begin{document}

\begin{titlepage}
\setcounter{page}{1}
\makeheadline

\title {A new X-ray mission to measure the power spectrum of
fluctuations in the Universe}

\author{{\sc X. Barcons}, Santander, Spain \\
\medskip
{\small Instituto de F\'{\i}sica de Cantabria (Consejo Superior de
Investigaciones Cient\'{\i}ficas - Universidad de Cantabria)} \\
\bigskip
{\sc A.C. Fabian}, Cambridge, UK\\
\medskip
{\small Institute of Astronomy} \\
%\bigskip
%{\sc C.o. Author2}, City2, Country2\\
%\medskip
%{\small City's gastronomical institute} \\
}

\date{Received; accepted } 
\maketitle

%%%%%\summary
%%%%% Text of summaryEND
\summary
We propose a new, simple, dedicated X-ray mission to measure the power
spectrum of density fluctuations in the Universe, by accurately
mapping the X-ray background on the whole sky on scales of $\sim 1\, 
{\rm deg}^2$. Since the method relies on the detection of excess
fluctuations produced by the clustering of X-ray sources at
intermediate redshifts above the confusion noise produced by bright foreground
sources, counting noise and systematic noise need to be
kept to a minimum.  We propose a proportional counter
with collimated fields of view of various sizes as the optimal
instrument for this mission.
END

%%%%% \keyw
%%%%% Your list of keywords
%%%%% END
\keyw
The X-ray Background, Large Scale Structure of the Universe.
END

%%%%% Astronomy and Astrophysics Abstracts classification, type in if known
\AAAcla
END
\end{titlepage}

%
%%%%% Chapter: 
%%%%% \kap{Title of chapter}
%%%%% Section: 
%%%%% \sect{Title of section}
%%%%% Subsection: 
%%%%% \subsect{Title of subsection}
\kap{Introduction} 

Recent progress in the optical identification of faint X-ray sources
discovered in {\it ROSAT} deep fields shows that the bulk of
the soft (0.5-2~keV) X-ray background (XRB) originates at intrmediate redshifts
$z\approx 1-2$ (Boyle et al 1994, McHardy et al
1997, Hasinger et al 1997, Schmidt et al 1997). The X-ray volume
emissivity of the Universe in the soft band peaks at these
intermediate redshifts.  This is particularly interesting, since it is
at these redshifts where the process of galaxy formation culminates
its non-linear growth and where the bulk of star formation 
appears to happen (Madau et al 1996). Studies of the angular structure
of the XRB should be able to map the lumpiness of the Universe at
these redshifts.

The distribution of density fluctuations in the Universe on different
scales is described in terms of the power spectrum (PS) of the
fluctuations ${\cal P}(z,k_c)$ ($z$ is the redshift and $k_c$ the
comoving wavenumber), which is the Fourier transform of the
two-point correlation function.  The current information on the PS
comes from both local galaxy surveys and from studies of anisotropies
in the microwave background radiation (MBR).  The former are most powerful
on small scales $<50\, h^{-1}\, {\rm Mpc}$ ($h$ is the Hubble constant
in units of $100\, {\rm km}\, {\rm s}^{-1}\, {\rm Mpc}^{-1}$) and the
latter emphasise large scales ($\sim 1000\, h^{-1} {\rm Mpc}$) at very
high redshifts ($z\sim 1500$) when the amplitude of the PS was very
small. Overall fits of the PS can be reproduced by Cold Dark Matter
models where the PS presents a broad peak at  comoving wavevectors
$k_c\sim 0.01 - 0.1\,  h {\rm Mpc}^{-1}$ (see Peacock 1997 for recent
parametrisations).   

In this paper we propose a new X-ray mission dedicated to measure the
PS near its maximum at the redshifts where the bulk of the XRB
originates.  There are various reasons to study this specific region
of the redshift-wavevector space. The most important one is that, as
has been said, it provides a measurement at a redshift in
between the very smooth Universe at the MBR epoch and the very lumpy
Universe we see today. Moreover, the evolution of the PS between
$z\sim 2$ and $z=0$ is most sensitive to the specific cosmological
model ($q_0$ and $\Lambda$). And finally, by selecting angular scales
of $\sim 1$~deg, we are most sensitive to the peak to the PS, where
the fluctuations are easier to measure and most useful to define the
normalization of the PS.

The precision of intensity measurements of the extragalactic XRB on
these scales is dominated by spatial fluctuations caused by confusion
noise. Sources with 2-10~keV fluxes $\sim 10^{-12}\, {\rm erg}\, {\rm
cm}^{-2}\, {\rm s}^{-1}$ will dominate this confusion noise. The
measurement of excess fluctuations produced by clumping of the more
numerous, distant sources requires both a precise knowledge of the
confusion produced by the foreground sources and also precise
measurements of the XRB intensity over the whole sky. Source counts
down to a 2-10~keV flux level $< 10^{-12}\, {\rm erg}\, {\rm
cm}^{-2}\, {\rm s}^{-1}$ will be found in the all-sky survey performed
by {\it ABRIXAS} (Friedrich et al 1996). Source variability might be a
problem (see later) if large, due to the non-simultaneous
observations of the sources and the XRB measurements. Photon counting
noise might be a limiting factor in the precision of the XRB
measurements, and therefore a large effective area is needed.  The
requirement that other systematics are kept to a minimum calls for the
most stable instrumentation, large effective area and minimal
degradation and gain variations over long periods.  In what follows,
we outline the mission requirements in terms of measurement requirements
for cosmological purposes and then study with some detail
the case of a collimated field-of-view proportional counter, which is
the instrument we propose. We concentrate on the 2-10~keV energy band
where the contribution from the Galaxy at high galactic latitude is
likely to be relatively small and smooth on scales of $\sim 1$~deg.

\kap{Sensitivity to excess fluctuations}

The distribution of XRB intensities when measured with a beam of
$\Omega\, {\rm deg}^2$ is a non-gaussian `P(D)' curve with various
contributions to its intrinsic dispersion $\left( {\Delta I\over
I}\right)_{int}$.  We explore these contributions in the next section
(confusion noise, photon counting noise and systematics) and separate
here the excess fluctuations $\left( {\Delta I\over
I}\right)_{excess}$ that we intend to measure. These excess
fluctuations can be related to ${\cal P}(z,k_c)$ weighted by
the square of the X-ray volume emissivity ($j(z)$) with appropriate
K-corrections.  The volume emissivity $j(z)$ is still not known for
the 2-10~keV band, but it certainly will be after {\it AXAF} deep surveys
have been performed (by resolving virtually 100\% of the XRB) and
subsequent optical follow-up with 10-m class telescopes has revealed
the redshift distribution of the sources that give rise to the XRB.
For the moment we assume that $j(z)$ is similar to the 0.5-2~keV one,
which peaks strongly at a redshift $z\sim 1.5-2$ (say
$z_c=1.7$). Roughly speaking, the expected excess fluctuations are 
\[
\left( {\Delta I\over I}\right)_{excess}^2={2\over \Delta V} {\cal
P}(z_c,k_0)
\]
where $\Delta V$ is the volume of space sampled by a single beam
$\Omega$ and $k_0\approx 0.04\Omega^{-1/2}\, h\, {\rm Mpc}^{-1}$ is
the wavevector at which the angular selection function peaks (see
Barcons, Fabian \& Carrera 1997 for details).  This is very close to
the expected peak of the PS, which has the added advantage that small
changes in the angular scale produce a negligible change in the value
of the PS.  To make a significant detection (see Fig.~5 in Barcons,
Fabian \& Carrera 1997) a sensitivity of less than $\sim 1 h^{-3}\,
{\rm Mpc}^3$ should be achieved in terms of the power spectrum.  The
volume sampled by a beam is of the order of $10^5\, \Omega\, {\rm
Mpc}^3$, and therefore excess fluctuations of less than $5\times
10^{-3}\Omega^{-1/2}\sim 0.5\%$ should be measured for a detection.

The minimum value of the excess fluctuations that can be measured when
$N_{obs}$ independent measurements of the XRB are performed is 
\[
\left( {\Delta I\over I}\right)_{2\sigma}=\sqrt{2\over N_{obs}}\left(
{\Delta I\over I}\right)_{int}
\]
As we shall see in next section, to achieve a value of a few $\times
10^{-3}$, this requires $N_{obs}\sim 10^4$, which means approximate all-sky
coverage for a $\sim 1$~deg$^2$ beam.

\kap{A collimated field of view proportional counter}

In this section we evaluate the budget of contributions to the
intrinsic dispersion of the distribution of XRB intensities, for a
collimated field of view proportional counter with effective area
$10^4 A_4\, {\rm cm}^2$, exposure time per pointing $100t_{100}\, {\rm s}$
and beamsize $\Omega\, {\rm deg}^2$. 

\sect{Confusion Noise}

Assuming an euclidean power-law for the bright source counts in the
2-10~keV band (Piccinotti et al 1982, Gendreau, Barcons \& Fabian
1997, Georgantopoulos et al 1997, Cagnoni, Della Ceca \& Maccacaro
1997), the contribution to the intrinsic dispersion from confusion
noise should be
\[
\left( {\Delta I\over I} \right)_{confusion}\sim 0.13 \Omega^{-1/3}
\]
Since our goal is to measure fluctuations almost 100 times smaller
than this, this number should be known with a few per cent accuracy.
The {\it ABRIXAS} all-sky survey will provide the required accuracy,
but two systematic effects should be carefully taken into account.

\subsect{Source Variability}

Variable sources produce a systematic effect in the determination of
source counts in flux-limited samples (Barcons, Fabian \& Carrera
1997). Assuming varability within a factor of 2, this bias is of the
order of 1\%, but for a factor of 5 variability this amounts to a 5\%
effect (assuming euclidean counts).  This effect can be corrected for
if known.  That should be possible in {\it ABRIXAS} at least for the
brightest sources, since the sky will be scanned several times.

\subsect{Spectrum of bright sources}

Since the spectral responses of the instrument proposed and {\it
ABRIXAS} (with which the source counts at bright fluxes will be found)
are different, there will be an indetermination when translating the
source counts, particularly with highly absorbed sources.  This could
be resolved by examining a sample of bright sources detected with our
instrument. A comparison with observations of {\it ABRIXAS}, should be
able to remove most of this uncertainty.

\sect{Photon counting Noise}

Events detected in a proportional counter come from both X-ray and
charged particles passing through the detector.  Since both fluxes
scale differently with the field-of-view solid angle, it is crucial to
have at least 2 different collimator sizes for accurate particle
background subtraction (Boldt 1987).  

The counting noise resulting from both components has been estimated
by using `clean' sequences from both the {\it Ginga} LAC and the {\it
RXTE} PCA, resulting in
\[
\left( {\Delta I\over I}\right)_{counting}\approx 0.024
(A_4t_100)^{-1/2}\left[ {1\over \Omega}+{1\over \Omega^2}\right]^{1/2}
\]

\sect{Systematics}

One of the key issues is the stability of the particle background.
Experience accumulated with {\it Beppo}SAX suggests that an equatorial
orbit is the best choice for this purpose.  As already said, having at
least two different collimator sizes will allow accurate subtraction
(as opposed to modelling) of the particle background.

Stochastic gain variations in the proportional counter are the most
unpredictable source of systematics in our study.  Long term drifts
can be monitored and modelled appropriately with the observations
themselves, if the product $A_4t_{100}$ is large enough. For
$A_4t_{100}$ around 1, the statistical accuracy in the measurement of
each measurement of the XRB towards each direction will be  of the order of
3\%. Long-term gain variations should be detectable to a precission
much better than this, but this requires a large enough collecting area. 

\kap{Conclusions}

\begin{table*}
\centering
\caption{Summary of requirements for the proposed mission}
\begin{tabular}{|l|l|}
\hline {\bf Instrument} &\\
\hline 
Gas-filled proportional counter & \\
Collimator FOV & $0.5\times 2\, {\rm deg}^2$\\
                      & $1\times 2\, {\rm deg}^2$\\
Effective area & $\sim 1000\, {\rm cm}^2$\\
Energy bandpass & $\sim 2-10$ keV\\
Accuracy in measuring XRB & $<3\%$\\
\hline 
{\bf Mission/Payload} &\\
\hline Expected lifetime & $> 2$ years\\
Stabilization & One axis \\
Orbit & Equatorial\\
\hline
\end{tabular}
\end{table*}

Our proposal for measuring the peak of the power spectrum of the
fluctuations in the Universe at intermediate redshifts and comoving
wavevectors $k_c\sim 0.01-0.1\, h\, {\rm Mpc}^{-1}$, consists of a
mission which would scan the whole sky several times (at least 4) with
a proportional counter.  At least two different collimated field of
view sizes should be available to give a clean sky signal.  To have a
maximally stable particle background, an equatorial orbit would be
most desirable. A large effective area (close to $10^4\, {\rm cm}^2$)
is required to model long-term gain variations.  Table 1 summarizes
the main requirements for this mission.

The proposed mission is tecnologically very simple.  Proportional
counters are the most tested and stable X-ray detectors. A slightly
modified version of the {\it Ginga} Large Area Proportional Counter,
with two different collimator sizes, would certainly be good enough
for our purposes.  Also the requirements for the payload would be
minimal, since no 3-axis stabilisation would be needed during the
all-sky scans.

\refer
\aba
\rf{Barcons X., Fabian A.C., Carrera F.J., 1997, MNRAS, in the press
(astro-ph/9705180)}
\rf{Boldt E., 1987, Phys. Rep., 146, 215}
\rf{Boyle, B.J., Shanks, T., Georgantopoulos, I., Stewart,
G.C., Griffiths, R.E.,  1994, MNRAS, 271, 639}
\rf{Cagnoni I., Della Ceca R., Maccacaro T., 1997, ApJ, in the press (astro-ph/9709018)}
\rf{ Friedrich, P., Hasinger, G., Richter, G., Kritze, K.,
Tr\"umper, J., Br\"auninger, H., Predehl, P., Staubert, R.,
Kendizorra, E., 1996, In: R\"ontgenstrahlung from the Universe,
eds. Zimmermann, H.U., Tr\"umper, J., Yorke, H.; MPE Report 263,
p. 681}
\rf{Gendreau K., Barcons X., Fabian A.C., 1997, MNRAS, in the press
(astro-ph/9711083)}  
\rf{Georgantopoulos, I., Stewart, G.C., Blair, A.J., Shanks,
T., Griffiths, R.E., Boyle, B.J., Almaini, O., Roche, N., 1997, MNRAS,
in the press}
\rf{Hasinger G., Burg R., Giacconi R., Schmidt M., Tr\"umper J.,
Zamorani G., 1997, A\&A, in the press (astro-ph/9709142)}
\rf{Madau P., Farguson H.C., Dickinson M.E., Giavalisco M., Steidel
C.C., Fruchter A., 1996, MNRAS, 283, 1388}
\rf{McHardy I.M., 1997, MNRAS, in the press (astro-ph/9703163)}
\rf{Peacock J.A., 1997, MNRAS, 284, 885}
\rf{Piccinotti G., Mushotzky, R.F., Boldt, E., Holt, S.S.,
Marshall, F.E., Serlemitsos, P.J., Shafer, R.A., 1992, ApJ, 253, 485}
\rf{Schmidt M., Hasinger G., Gunn J., Schneider D., Burg R., Giacconi
R., Lehmann I., MacKenty J., Tr\"umper J., Zamorani G., 1997, A\&A, in
the press (astro-ph/9709144)}

\abe

\acknowledgements
We thank Elihu Boldt, Francisco Carrera, Keith Jahoda, G\"unther
Hasinger, Ofer Lahav, Gordon Stewart and Bob Warwick for valuable
suggestions which substantially improved this proposal.  XB
acknowledges partial financial support for this research to the DGES
under project PB95-0122 and grant PR95-490.  ACF thanks the Royal
Society for support.

%
%%%%% Address of the authors
\addresses
\rf{X. Barcons, 
Instituto de F\'{\i}sica de Cantabria (Consejo Superior de
Investigaciones Cient\'{\i}ficas - Universidad de Cantabria),
Facultad de Ciencias, 
Av. Los Castros, 
39005 Santander, 
Spain, 
e-mail: barcons@ifca.unican.es}
\rf{A.C. Fabian, Institute of Astronomy, Madingley Road, Cambridge CB3
0HA, UK, e-mail: acf@ast.cam.ac.uk}
END
%
%%%%% End of address

\end{document}